\documentclass[5p,times]{elsarticle}

\usepackage{lineno,hyperref}
\modulolinenumbers[5]
\usepackage{tikz}

\definecolor{mygreen}{RGB}{0, 190, 0}

\usepackage{graphicx}
\usepackage{amsmath}
\usepackage{amssymb}

\DeclareMathOperator*{\argmax}{arg\,max}

\definecolor{mygreen}{RGB}{0, 190, 0}

\usepackage[scientific-notation=true]{siunitx}

\usepackage{hyperref} 

\usepackage[normalem]{ulem}

\usepackage{arydshln}

\usepackage{natbib}

\usepackage{url}
\makeatletter
\g@addto@macro{\UrlBreaks}{\UrlOrds}
\makeatother

\journal{Journal of \LaTeX\ Templates}

\biboptions{authoryear}

\hypersetup{draft}
\begin{document}

\begin{frontmatter}

\title{Multi-atlas image registration of clinical data with automated quality assessment using ventricle segmentation}

\author[jpk,bigr]{Florian~Dubost
\corref{mycorrespondingauthor}}
\cortext[mycorrespondingauthor]{Corresponding author}
\ead{floriandubost1@gmail.com}

\author[bigr,dtu]{Marleen~de~Bruijne}
\author[jpk]{Marco~Nardin}
\author[csail]{Adrian~V.~Dalca}
\author[jpk]{Kathleen~L.~Donahue}
\author[jpk]{Anne-Katrin~Giese}
\author[jpk]{Mark~R.~Etherton}
\author[martino]{Ona~Wu}
\author[bigr,epi]{Marius~de~Groot}
\author[bigr,tudelft]{Wiro~Niessen}
\author[radio,epi]{Meike~Vernooij}
\author[jpk]{Natalia~S.~Rost}
\author[jpk,csail,dzne]{Markus~D.~Schirmer}
\ead{mschirmer1@mgh.harvard.edu}
\author{for the Alzheimer’s Disease Neuroimaging Initiative\fnref{fn1}}
 
\address[jpk]{J. Philip Kistler Stroke Research Center, Department of Neurology, Massachusetts General Hospital, Harvard Medical School, Boston, USA}
\address[bigr]{Biomedical Imaging Group Rotterdam, Department of Radiology and Nuclear Medicine, Erasmus MC - University Medical Center Rotterdam, The Netherlands}
\address[dtu]{Department of Computer Science, University of Copenhagen, Copenhagen, Denmark}
\address[csail]{Computer Science and Artificial Intelligence Lab, Massachusetts Institute of Technology, Cambridge, USA}
\address[martino]{Athinoula A. Martinos Center for Biomedical Imaging, Department of Radiology, Massachusetts General Hospital, Charlestown, MA, USA}
\address[epi]{Department of Epidemiology, Erasmus MC - University Medical Center Rotterdam, the Netherlands}
\address[tudelft]{Department of Imaging Physics, Faculty of Applied Science, TU Delft, Delft, The Netherlands}
\address[radio]{Department of Radiology and Nuclear Medicine, Erasmus MC - University Medical Center Rotterdam, the Netherlands}
\address[dzne]{Department of Population Health Sciences, German Centre for Neurodegenerative Diseases (DZNE), Germany}

\fntext[fn1]{Data used in preparation of this article were obtained from the Alzheimer’s Disease Neuroimaging Initiative (ADNI) database (adni.loni.usc.edu). As such, the investigators within the ADNI contributed to the design and implementation of ADNI and/or provided data but did not participate in analysis or writing of this report. A complete listing of ADNI investigators can be found at: \url{http://adni.loni.usc.edu/wp-content/uploads/how_to_apply/ADNI_Acknowledgement_List.pdf}}

\begin{abstract}
Registration is a core component of many imaging pipelines. In case of clinical scans, with lower resolution and sometimes substantial motion artifacts, registration can produce poor results. Visual assessment of registration quality in large clinical datasets is inefficient. In this work, we propose to automatically assess the quality of registration to an atlas in clinical FLAIR MRI scans of the brain. The method consists of automatically segmenting the ventricles of a given scan using a neural network, and comparing the segmentation to the atlas’ ventricles propagated to image space. We used the proposed method to improve clinical image registration to a general atlas by computing multiple registrations - one directly to the general atlas and others via different age-specific atlases - and then selecting the registration that yielded the highest ventricle overlap. Finally, as an example application of the complete pipeline, a voxelwise map of white matter hyperintensity burden was computed using only the scans with registration quality above a predefined threshold. Methods were evaluated in a single-site dataset of more than 1000 scans, as well as a multi-center dataset comprising 142 clinical scans from 12 sites. 
The automated ventricle segmentation reached a Dice coefficient with manual annotations of 0.89 in the single-site dataset, and 0.83 in the multi-center dataset. Registration via age-specific atlases could improve ventricle overlap compared to a direct registration to the general atlas (Dice similarity coefficient increase up to 0.15). Experiments also showed that selecting scans with the registration quality assessment method could improve the quality of average maps of white matter hyperintensity burden, instead of using all scans for the computation of the white matter hyperintensity map.
In this work, we demonstrated the utility of an automated tool for assessing image registration quality in clinical scans. This image quality assessment step could ultimately assist in the translation of automated neuroimaging pipelines to the clinic.
\end{abstract}

\begin{keyword}
Registration, ventricles, segmentation, deep learning, quality, multi-atlas, age, white matter hyperintensity, ADNI
\end{keyword}

\end{frontmatter}

\section{Introduction}

Image registration has proven a fundamental part of many processing pipelines in the biomedical imaging field, establishing spatial correspondence between images and enabling subsequent group or cohort analyses. However, when using clinical, low resolution brain data, image registration can be challenging. E.g. in acute ischemic stroke populations, high-resolution image acquisition in the acute disease state is not possible due to clinical time constraints. Nonetheless, such clinical cohorts offer great amounts of untapped information due to the large number of samples available, often in the range of thousands of patients \citep{giese2017, courand2019}, which can be utilized to unveil spatial patterns of disease burden \citep{bilello2016,schirmer2019spatial}. Importantly, as clinical images have more variability than scans acquired primarily for research, they necessitate quality control steps after registration to ensure that no gross errors occurred in the process. Quantifying the registration quality, utilizing only intensity-based metrics such as mutual information or cross-correlation, is often not enough, and in practice registration quality is assessed using manual ventricle segmentations to evaluate the overlap between the patient data and the registration target, i.e. brain template or atlas \citep{ou2014,dalca2016,ganzetti2018}. 

Considerable work has been conducted to generate appropriate brain templates for image registration, using data from healthy young adults \citep{dickie2017} or age appropriate cohorts from the general population \citep{schirmer2019spatial}. These templates can consequently be used for segmentation of brain structures, but often yields unsatisfactory results in clinical scans. For instance, outlining of the ventricles in such clinical scans is often done manually, or semi-automatically \citep{hussain2013,xia2004}. Manually outlining the ventricles is a time intensive step, and hinders quality assessment in large scale cohorts. Deep learning techniques have been developed to automatically segment structures in clinical quality scans, using for instance U-Net architectures \citep{schirmer2019white,nikolov2018,guerrero2018}. Given enough training data, these techniques can reliably generate accurate, fully automated masks of the structures of interest. The use of a U-Net architecture has been proposed to generate automated segmentations of the lateral ventricles alone \citep{ghafoorian2018}, and recently of the complete ventricular system \citep{atlason2019, shao2019}, showing promising results, which can be utilized in automated assessment of image registration quality. 

Automated registration quality assessment methods can also be used to improve the registration results in atlas selection methods. Multi-atlas segmentation has for instance become an increasingly popular segmentation method in neuroimaging pipelines \citep{iglesias2015}. One of its simplest implementations is to register several atlases pairwise to an image, propagate the labels of the atlases in image space, and choose the final label for each voxel using majority voting. Probabilistic label fusion strategies have also been proposed, such as \cite{wang2013} who proposed to exploit the intensity similarity between atlases and the target image in the neighborhood of each voxel. \cite{robinson2019} recently proposed a method to perform automated quality control of segmentations of cardiovascular data from the UK biobank. The authors registered a set of annotated images to a test image with unknown ground truth. The labels were then warped using the deformation field from image registration, and the overlap between the warped labels and the predicted segmentation was used to estimated the segmentation performance. In other words, the segmentation of the image with unknown ground truth is compared to that of a multi-atlas segmentation, where smaller difference between segmentations are assumed to reflect higher segmentation quality.
Instead of using the same set of atlases for multi-atlas segmentation, a most appropriate subset of atlases can also be selected. Recently, \cite{antonelli2019} proposed for instance to select subsets of atlases for each target image using a genetic selection algorithm, and evaluated their method in cardiac and prostate data. To decrease the computation time of multi-atlas segmentation, \cite{dewey2017} proposed to add an intermediary registration step to a template constructed from the set of the considered atlases, using for instance multivariate template construction algorithm. Creating robust registration methods to map clinical scans to atlases is key to the field of lesion-symptom mapping. For example, \cite{biesbroek2013} studied lesion-symptom mapping with brain lesions, such as white matter hyperintensities and lacunes, in relation to cognition.

In this work, we developed a ventricle segmentation deep learning algorithm based on a 3D U-Net-like architecture to segment the complete ventricular system in each subject’s fluid-attenuated inversion recovery (FLAIR) sequence and validated it in a multi-center, clinical dataset comprising 12 sites. The ventricle segmentation was then used to assess registration quality by comparing it -- using the Dice similarity coefficient -- to the ventricles of the atlas propagated to the target image space. Over all brain regions, due to its very discriminative image intensity values and its relatively large size, the ventricular system presents a feature of the brain that is robust to variations in scanners and FLAIR protocols, making it a prime candidate for using its segmentation to assess registration quality. This automated registration quality assessment method can be used not only to flag or discard erroneous registrations, but also to select the best registration. As an example, we proposed to use this automated registration quality assessment method to improve registration quality by designing a multi-atlas registration (MAR) framework. Instead of directly registering images to a single template (general atlas), each image was additionally registered to five different atlases corresponding to different age categories, which in turn have been registered to the general atlas. The best atlas was then selected using the automated registration quality assessment method, and used as a transitional registration step before warping the subject image to the common space. Contrary to the above-mentioned multi-atlas segmentation methods, the purpose of the proposed MAR method was to improve the results of registration to the common space, and not to improve the results of segmentation of brain regions in the target image. Finally, we used the proposed MAR framework to create voxelwise maps of white matter hyperintensity (WMH) burden in a set of acute ischemic stroke patients, where Dice coefficient thresholds were used to control the quality of registration. In summary, our main contributions are an algorithm for the segmentation of the complete ventricular system in clinical scans, the evaluation of ventricle overlap as registration quality metric, and a multi-atlas registration framework to improve registration of images to a common space.

\begin{figure*}[!t]
\centering
\includegraphics[height=6cm]{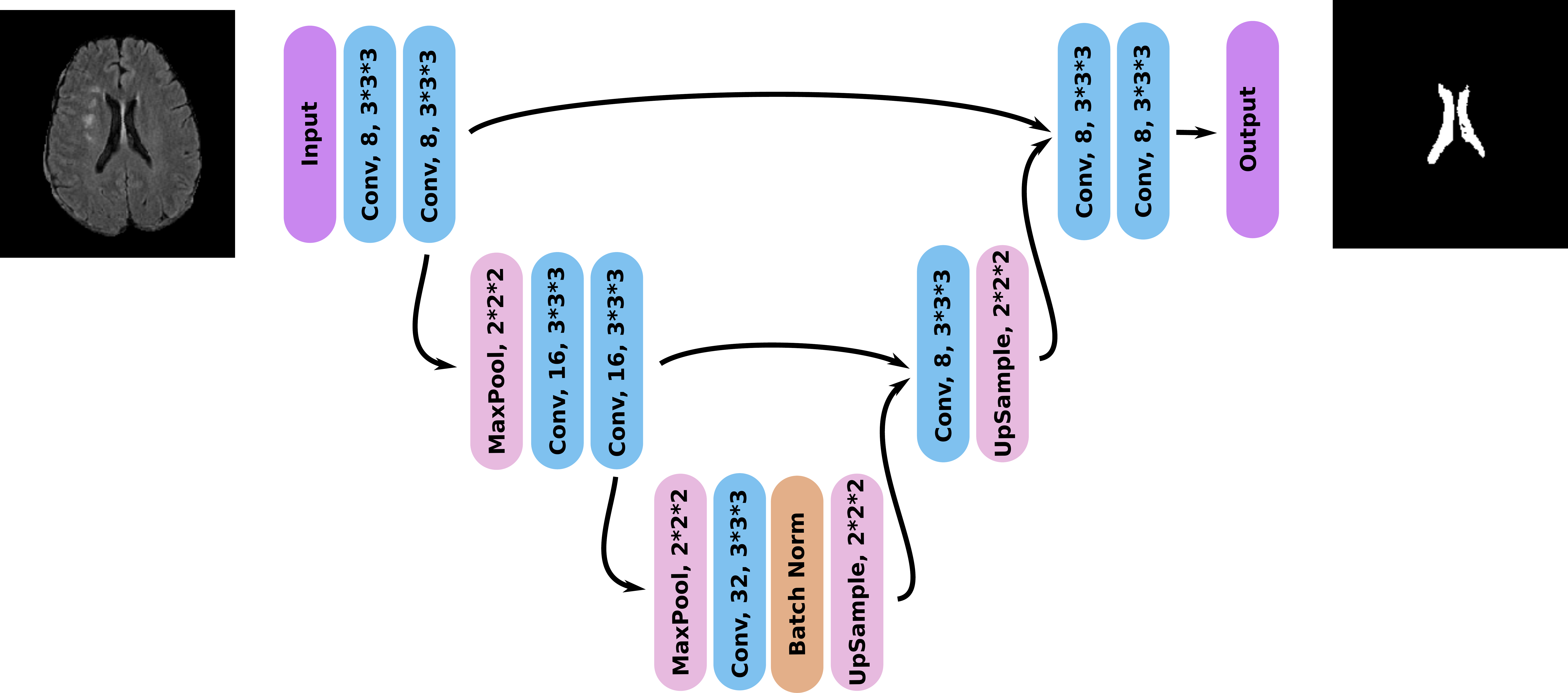}
\caption{\textbf{Architecture of the deep learning ventricle segmentation algorithm.} The architecture is similar to that of a shallow 3D U-Net \citep{ronneberger2015} with only 104 feature maps to allow the processing of the full 3D images.
}
\label{fig:arch}
\end{figure*}

\section{Material and Methods}

\subsection{Data}
\subsubsection{Onsite clinical data}
We utilized data of the Genes Affecting Stroke Risk and Outcomes Study (GASROS) study \citep{zhang2015}. Patients ($>18$ years old) presenting to the Massachusetts General Hospital Emergency Department (ED) between 2003 and 2011 with symptoms of acute ischemic stroke, were eligible for enrollment. Magnetic resonance images were acquired within 48 hours of admission and only patients with confirmed acute diffusion-weighted imaging lesions on brain MRI scans were included. 1132 patients underwent the standard acute ischemic stroke protocol on a 1.5T Signa scanner (GE Medical Systems), including T2-weighted FLAIR imaging (TR 5000ms, minimum TE of 62 to 116ms, TI 2200ms, FOV 220-240mm). For each patient, WMH were segmented using MRIcro software (University of Nottingham School of Psychology, Nottingham, UK; www.mricro.com), based on a previously published semi-automated method with high inter-rater reliability \citep{chen2006}. Ventricles were manually segmented by a single rater in a subset of 300 patients’ FLAIR images using 3D Slicer \citep{fedorov2012}. Of the 300 scans, 100 were chosen to uniformly sample the age range in the GASROS cohort, 100 were chosen to span the range of WMH disease burden, and the remaining 100 were randomly selected. This set was used for network training and validation of the automated ventricle segmentation method. In addition, a test set of 100 patients were selected to approximately represent the range of ventricular volume in the patient population. Scans were selected with a semi-automated method that estimates ventricular volume using nonlinear registration to an atlas. The semi-automated method involved a quality control step to ensure that the range was uniformly sampled. These 100 scans were then segmented by a second rater. 

\subsubsection{Multi-center clinical data}
The MRI-GENetics Interface Exploration (MRI-GENIE) study is a large-scale, international, hospital-based collaborative study of acute ischemic stroke patients \citep{giese2017}, including FLAIR data from 12 sites (7 European, 5 US based), acquired as part of each hospital’s clinical acute ischemic stroke protocol. For each acquisition site, 12 patients were selected \citep{schirmer2019white} and underwent manual ventricle segmentations. Two of the patients displayed substantial motion artifacts, and were excluded from our analysis, forming a total set of N=142 scans with manual brain and ventricle segmentation. This set was used as an additional test set for the evaluation of the ventricle segmentation algorithm and the proposed MAR framework.

\subsubsection{ADNI data}
Part of the data used in the preparation of this article were also obtained from the Alzheimer’s Disease Neuroimaging Initiative (ADNI) database (adni.loni.usc.edu). The ADNI was launched in 2003 as a public-private partnership, led by Principal Investigator Michael W. Weiner, MD. The primary goal of ADNI has been to test whether serial magnetic resonance imaging (MRI), positron emission tomography (PET), other biological markers, and clinical and neuropsychological assessment can be combined to measure the progression of mild cognitive impairment (MCI) and early Alzheimer’s disease (AD).

\subsubsection{Brain atlases}
Using 130 healthy controls from ADNI3 dataset \citep{jack2008} (Field strength 3T; 3D FLAIR; TE 119; TR 4800; TI 1650; 1.2x1x1mm3; see \ref{appendix:ADNIdata} for list of subject IDs), we created five FLAIR atlases, each corresponding to a different age category: under 70 years old (N=6 subjects), between 70 and 75 (N=22), between 75 and 80 (N=31), between 80 and 85 (N=39), and above 85 (N=32). The atlases were created using ANTs multivariate template construction algorithm with default parameters \citep{avants2011}. Similarly, a general atlas was created by averaging the five age-specific atlases, also using using ANTs multivariate template construction algorithm with default parameters \citep{avants2011}. All atlases were manually skull stripped and registered to MNI space. The resulting image resolution was 1mm3 and the image size 182x218x182 voxels. Ventricles were manually segmented in the general atlas. Each of the five age-specific atlases was diffeomorphically registered to the general atlas, to allow the propagation of the ventricle segmentation to age-specific atlases, and to warp the images to the general atlas space in the MAR framework. To assess which atlases were most similar to the general atlas, we computed the mean squared intensity difference between the age-specific atlases and the general atlas.

\begin{figure*}[!t]
\centering
\includegraphics[height=5cm]{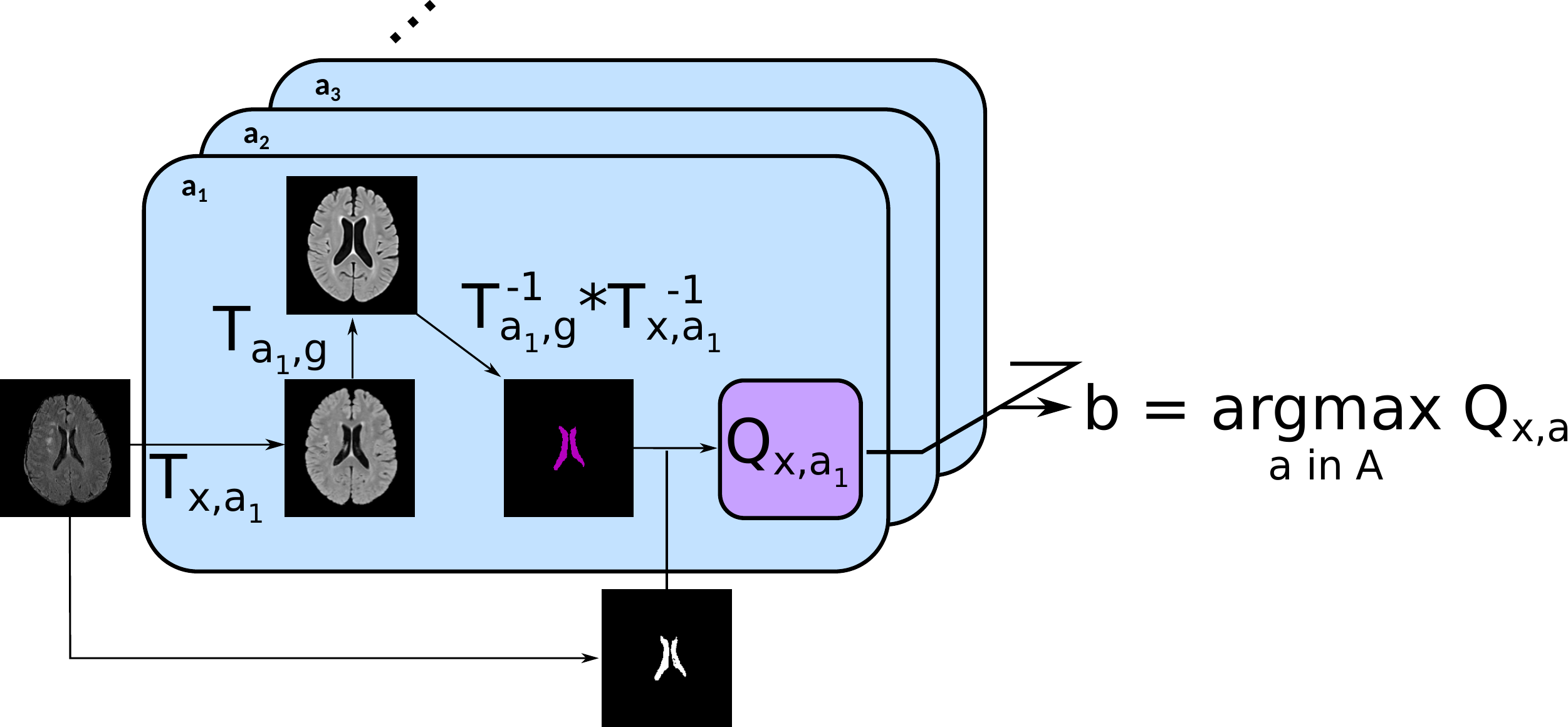}
\caption{\textbf{Principle of the proposed MAR framework.} For each subject, the input image was first registered to each of the atlases $a \in A$, which had been previously registered to the general atlas. The ventricles segmented on the general atlas $V_g$ are then propagated first to each atlas a, and then to the subject's image space. The propagated ventricles $V_{x,a,g}$ were subsequently compared to $V_{CNN}$, the subject's ventricles segmented using the proposed automatic algorithm. Finally, the atlas maximizing the registration quality was selected for the intermediary registration step.  
}
\label{fig:mar}
\end{figure*}

\subsection{Automated ventricle segmentation}
Image intensities were rescaled so that the 1st percentile of intensity values (without masking) is equal to 0 and the 99th percentile is equal to 1. The full 3D images were passed as input to a deep learning model. Prior to ventricle segmentation, each FLAIR image underwent brain extraction using a dedicated U-Net based deep learning method \citep{schirmer2019white} developed and validated in clinical scans. The resulting brain mask was also given as input to the model. While test data had varying voxel dimensions, training data consisted only of images with image size of 256x256 voxels in axial (inplane) direction, and less than 32 voxels in through plane direction. All images were then padded in z to have 32 slices. During inference, we resized images to 256x256x32 voxels using linear interpolation, predicted the corresponding ventricle maps, and resized these maps to the original image resolution.

We used a 3D U-Net-like architecture (Figure \ref{fig:arch}), based on two up-/down-sampling layers. Each convolution layer had a kernel size of 3x3x3 with ReLu activations, and we utilized 2x2x2 Max-Pooling for downsampling. To accelerate convergence without overloading the GPU memory, we added a Batch Normalization layer \citep{ioffe2015} after the features maps with the lowest resolution (5th convolution layer). Additionally, to improve generalization, we added a Dropout layer \citep{srivastava2014} before the last convolution. The parameters of the network were optimized with the Adadelta optimizer \citep{zeiler2012}. To improve generalisation, we also trained the algorithm with online data augmentation using random translations $<50$ voxels, 3D rotations of maximum 0.2 radian and flipping according to the coronal plane. The intensity of the ventricles and of the sulci were also separately randomized for data augmentation. To artificially increase the intensity of the ventricles, we used the annotations and randomly added to the ventricles intensities a maximum of $2\mu$, with $\mu$ the mean intensity of the FLAIR scans after percentile normalization. To artificially modify the intensity of the sulci, we randomly added between $-2\mu$ and $2\mu$ to regions of the images with an intensity value lower than 0.25 after percentile normalization. The algorithm was implemented using the publicly available Keras 2.2.0 library \citep{chollet2015} with TensorFlow 1.10 as backend \citep{abadi2016}.

The network’s outputs were binarized at a threshold of 0.5. To improve the segmentation, in the ventricle binary maps, we removed small connected components with a volume smaller than a manually determined threshold of 5 voxels.

\subsection{Registration quality assessment}

All pairwise registrations from image to atlas were performed using ANTs SyN nonlinear diffeomorphic registration algorithm with default parameters \citep{avants2011}. Inverse registrations were computed to allow the propagation of atlases’ ventricle segmentations to image space. The quality of the registration $T_{x,a}$ of an image $x$ to an atlas $a$ can be assessed by measuring the overlap between the ventricles segmented by the CNN in image space ($V_{CNN}$) and the ventricles of the atlas a ($V_a$) propagated to image space $V_{x,a}= T^{-1}_{x,a}(V_a)$. We denote this registration quality metric as $Q_{x,a} = D(V_{CNN},V_{x,a})$, where $D$ is the Dice similarity coefficient.

Other more conventional metrics -- that measure e.g. image similarity -- could be used instead to assess registration quality. We assessed this based on the cross-correlation (CC), i.e. the registration metric itself (ANTS SyN \citep{avants2008,sarvaiya2009,deGroot2013}) between the registered image $x$ and each atlas $a$ such that $Q_{x,a} = T_{x,a}(x) \star a$, where $\star$ denotes the cross-correlation operation. Prior to the computation of the cross-correlation, images were rescaled in $[0,1]$ using their minimum and maximum intensity values.

\subsection{Multi-Atlas Registration}
Each scan was registered pairwise to each atlas in $A={a_1,...,a_5,g}$, where $a_i$ are the age-specific atlases and the $g$ is the general atlas. For a given scan, the best atlas $b$ was then selected based on the registration quality metric $Q$, so that 

\begin{equation}
b = \argmax_{a \in A} Q_{x,a}, 
\end{equation}

with, for the ventricle overlap quality metric, $Q_{x,a}=D(V_{CNN}, V_{x,a,g})$, where $V_{x,a,g} =  T^{-1}_{x,a}T^{-1}_{a,g}(V_g)$. If the best atlas was not the general atlas, the scan uses the intermediate registration target $b$ and is then warped to the general atlas using the deformation field of the registration of the intermediary atlas to the general atlas (Figure \ref{fig:mar}).

\section{Experiments}
\subsection{Ventricle Segmentation}

The ventricle segmentation algorithm was optimized using the training/validation set, which was randomly split into 240 training scans and 60 validation scans to monitor over-fitting. The algorithm was then evaluated on the test set of 100 scans. The experiments with the MAR framework were conducted using the complete GASROS dataset excluding the 300 scans used to optimized the ventricle segmentation algorithm and 41 scans with strong motion artifacts, but excluding the 100 scans of the test set for ventricle segmentation, hence resulting in 791 scans.

We assessed the automatic segmentation of the ventricular system in the FLAIR sequences based on 11 different metrics. These metrics included the Dice similarity coefficient (Dice), Jaccard index (Jaccard), true positive rate (TPR), mutual information (MI), Cohen's kappa (KAP), intraclass correlation coefficient (ICC), volumetric similarity (VS), adjusted Rand index (ARI), probabilistic distance (PBD), detection error rate (DER) and outline error rate (OER).
VS was computed as the absolute volume difference divided by the sum of both volumes. ARI is Rand index corrected for chance. Rand index measures similarity between clusters. PDB measures the distance between fuzzy segmentations. DER measures the disagreement in detecting the same regions, namely the sum of the volumes of regions detected in only one of both segmentations. OER measures the disagreement in outlining of the regions, namely the difference between union and intersection of regions detected in both segmentations. A detailed description of the metrics is given elsewhere \citep{taha2015,wack2012}. 

PBD, DER, and OER are a measure of dissimilarity, where smaller values represent better agreement. As DER and OER are bounded metrics, we rescaled them between 0 and 1, and reported 1-DER and 1-OER. In case of PBD (not bounded), we reported 1/(1+PBD). Subsequently, all similarity metrics are bound between 0 and 1, where 1 indicates a perfect segmentation. Results are visualized as radar plots.\footnote{Github link - https://github.com/marconardin/spider-plotting}

\subsection{Evaluation of the multi-atlas registration framework}
We compared the proposed multi-atlas registration method to a direct registration to the general atlas and quantified the gain in registration performance by the difference $\Delta_{b,g} = Q_{x,b} - Q_{x,g}$, where $Q$ represents the Dice coefficient of ventricle overlap. We computed Wilcoxon tests on all subjects, in order to evaluate the efficacy of the proposed MAR framework. Additionally, we investigated the effect of utilizing different registration quality assessment metrics and the dependency of age and ventricle volume on the selection of the best atlas. 

\subsection{Spatial maps of WMH burden}
Utilizing the manual WMH segmentations from GASROS, we generated an average voxelwise map of WMH burden in template space. After using the MAR framework, we selected subjects for which registration quality was above a threshold $T$. Using three different thresholds $T$ = 0, 0.6, and 0.9, we visually assessed the quality of WMH maps constructed.

\section{Results}
\subsection{Ventricle segmentation}
\begin{figure*}[!t]
\centering
\includegraphics[height=9cm]{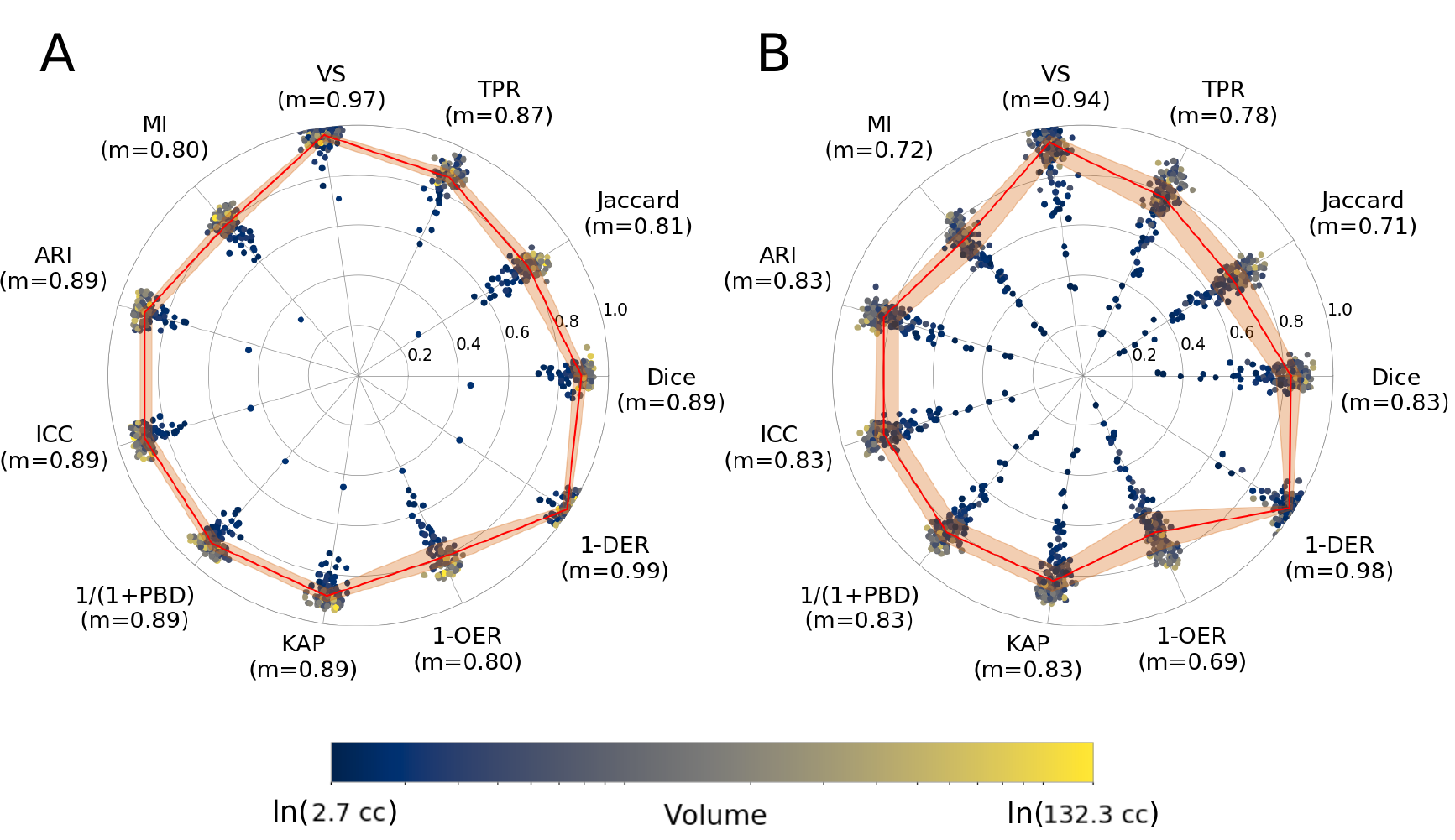}
\caption{\textbf{Comparison of automated and manual ventricle segmentations in A) GASROS (N=100; left) and B) MRI-GENIE (N=142; right).} The reported metrics are Dice coefficient (Dice), Jaccard index (Jaccard), true positive rate (TPR), volumetric similarity (VS), Mutual information (MI), Adjusted Rand Index (ARI), intraclass correlation coefficient (ICC), probabilistic distance (PBD), Cohen’s kappa (KAP), Detection Error Rate (DER) and Outline Error Rate (OER). The solid line is based on the median of each measure, while the ribbon represents the interquartile range.}
\label{fig:ventRes}
\end{figure*}

The results of evaluating the automated ventricle segmentation (see Figure \ref{fig:ventRes}) show good agreement between the manual and automated ventricle segmentations, with Dice coefficients of 0.89 for the single-site GASROS dataset and 0.83 for the multi-site MRI-GENIE dataset. Results of the ventricle segmentation for the MRI-GENIE data set, stratified by site, are shown in \ref{appendix:ventSeg}.

\subsection{Multi-atlas registration}
\subsubsection{Atlas creation}
\begin{figure*}[!t]
\centering
\includegraphics[height=3.5cm]{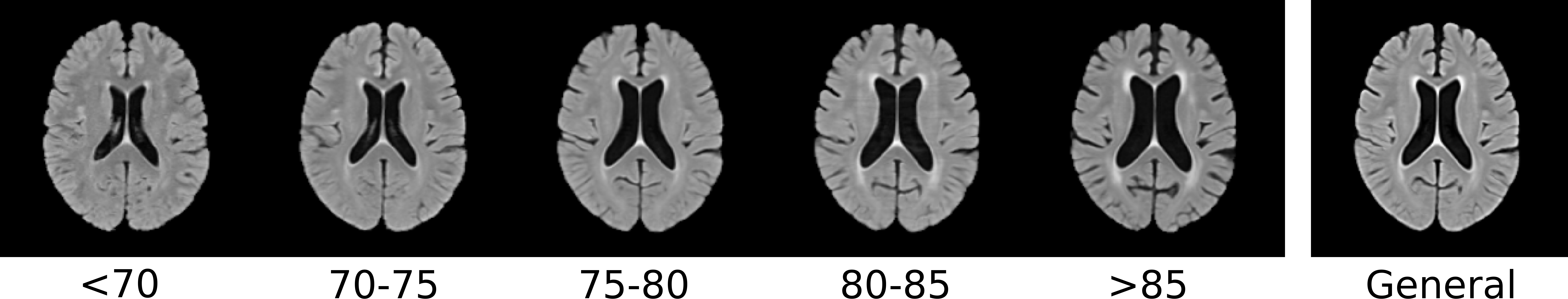}
\caption{\textbf{Age-specific atlases and the general atlas registered to MNI space.}}
\label{fig:atlases}
\end{figure*}

Figure \ref{fig:atlases} shows the age-specific atlases created from the healthy controls from the ADNI dataset. Computing the mean squared intensity difference between the age-specific atlases and the general atlas revealed that atlas 75-80 was the closest to the general atlas, and atlas 80-85 was the most dissimilar.

\subsubsection{Gain in registration performance}
\begin{figure*}[!t]
\centering
\includegraphics[height=7cm]{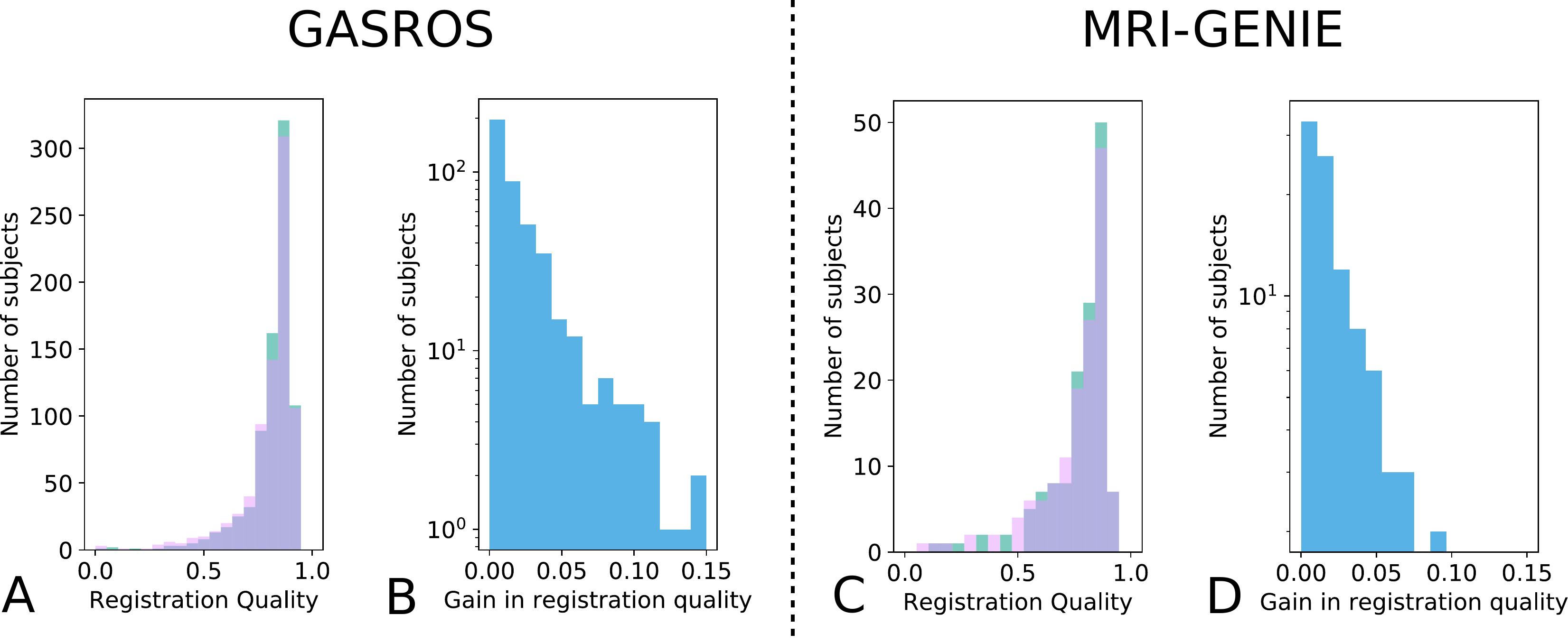}
\caption{\textbf{Gain in registration performance measured as ventricle overlap by using the proposed MAR method in comparison to a direct pairwise registration to the general atlas g for each dataset (Left: GASROS; Right: MRI-GENIE).} A/C: registration quality histograms using either direct registration to the general atlas (pink) or the MAR (green; improvement of registration quality). The overlap of both methods is shown in purple. B/D: Gain in registration quality $\Delta_{b,g}$. Scatterplots are also available in Appendix \ref{appendix:scatterplotDICE}.}
\label{fig:gain}
\end{figure*}

\begin{figure*}[!t]
\centering
\includegraphics[height=7cm]{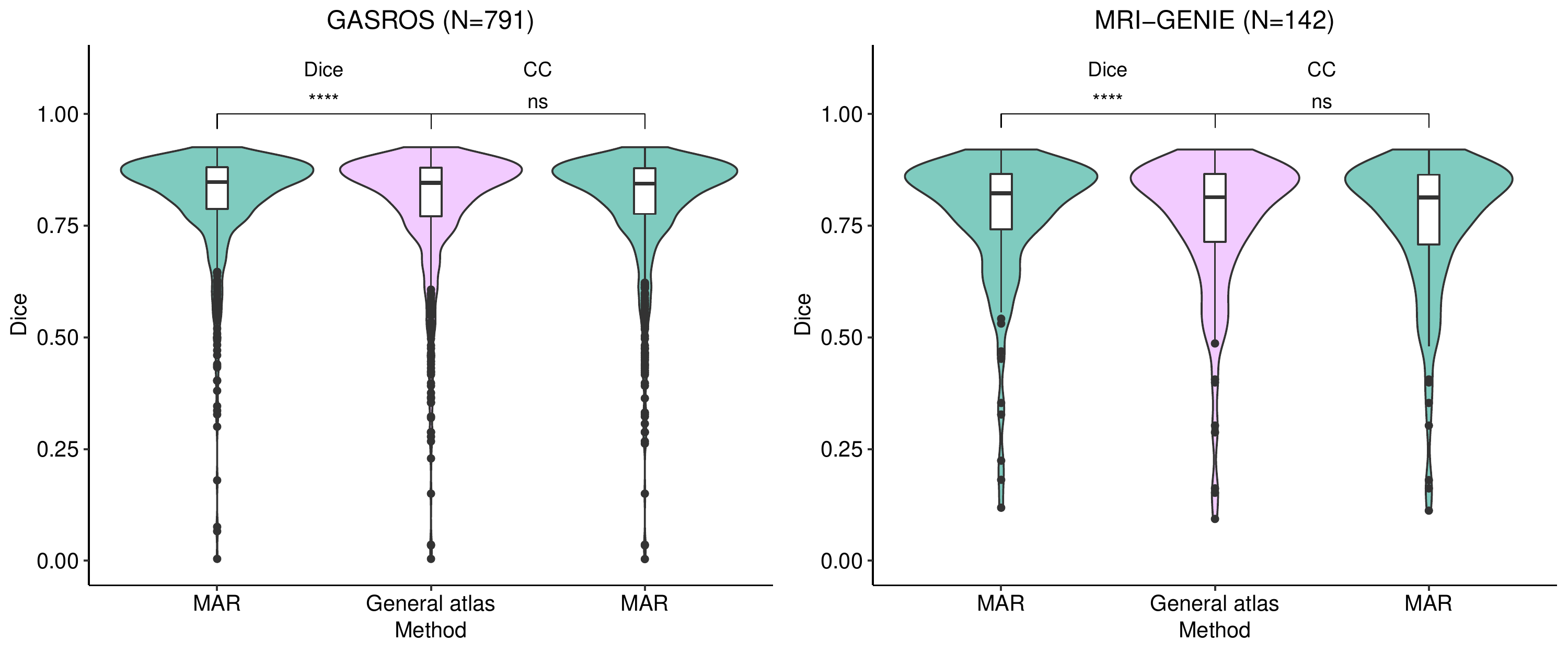}
\caption{\textbf{Comparison of the proposed MAR with a direct registration to the general atlas.} Instead of the proposed selection strategy for the intermediary atlas (ventricle Dice), we also experimented using the more standard selection criterion: cross-correlation (CC), computed after the elastic registration and normalization of intensity values. **** indicates a p-value lower than 0.0001 for the Wilcoxon test, and n.s. Indicates a non significant difference. }
\label{fig:marStat}
\end{figure*}

\begin{figure*}[!t]
\centering
\includegraphics[height=5cm]{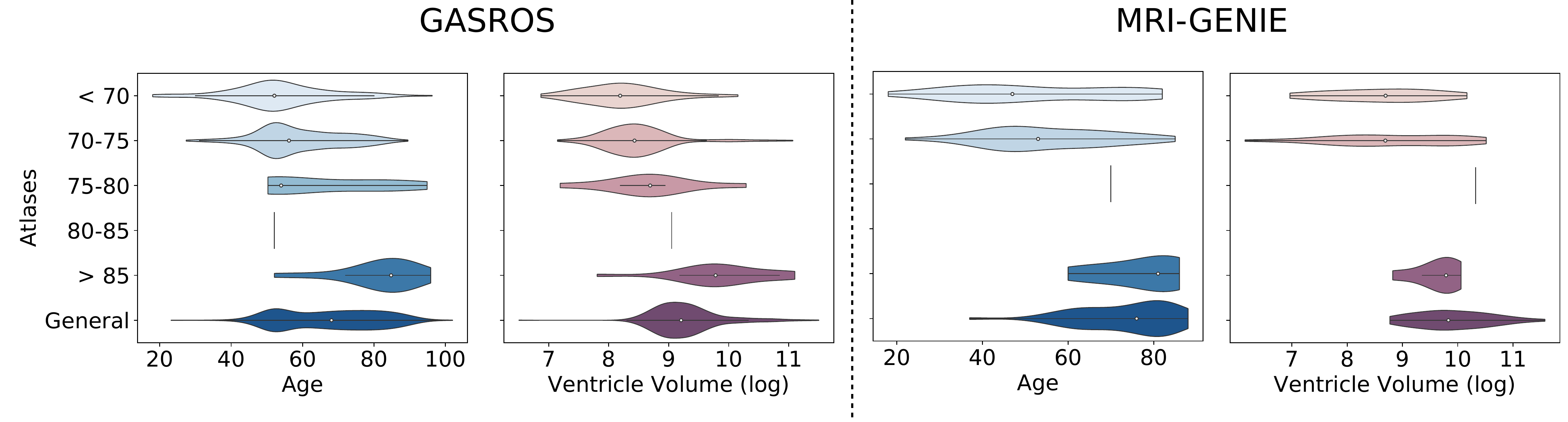}
\caption{\textbf{ Effect of age and ventricle volume on the selection of the atlases using ventricle overlap as registration quality metric.} Violin plots show the distribution of the subjects' age -- and ventricle volume -- according to the best atlas the subjects were assigned to in the MAR framework. A vertical line indicates that only n=1 subject has been assigned to the template.}
\label{fig:effectAge}
\end{figure*}

The gain in registration performance $\Delta_{b,g}$ is shown for each dataset in Figure \ref{fig:gain} and \ref{appendix:gainMAR}. We observed age-dependent improvements with increases of ventricle overlap by up to 0.15 Dice points. Wilcoxon tests showed that the proposed MAR method reached a significantly higher registration quality -- measured as ventricle overlap -- than that of the direct registration to the general atlas (Figure \ref{fig:marStat}) in N=430 GASROS subjects ($54\%$) and 93 MRI-GENIE subjects ($65\%$). However, when using cross-correlation instead of ventricles overlap for intermediary atlas selection, the proposed MAR method did not reach a significantly higher registration quality than that of the direct registration to the general atlas (Figure \ref{fig:marStat}; \ref{appendix:gainCCplot} and \ref{appendix:gainCCtable}). As expected, younger patients with lower ventricle volume were assigned to atlases of younger categories (Figure \ref{fig:effectAge}).

\subsubsection{Manual versus automated ventricle segmentation}
\begin{figure*}[!t]
\centering
\includegraphics[height=6cm]{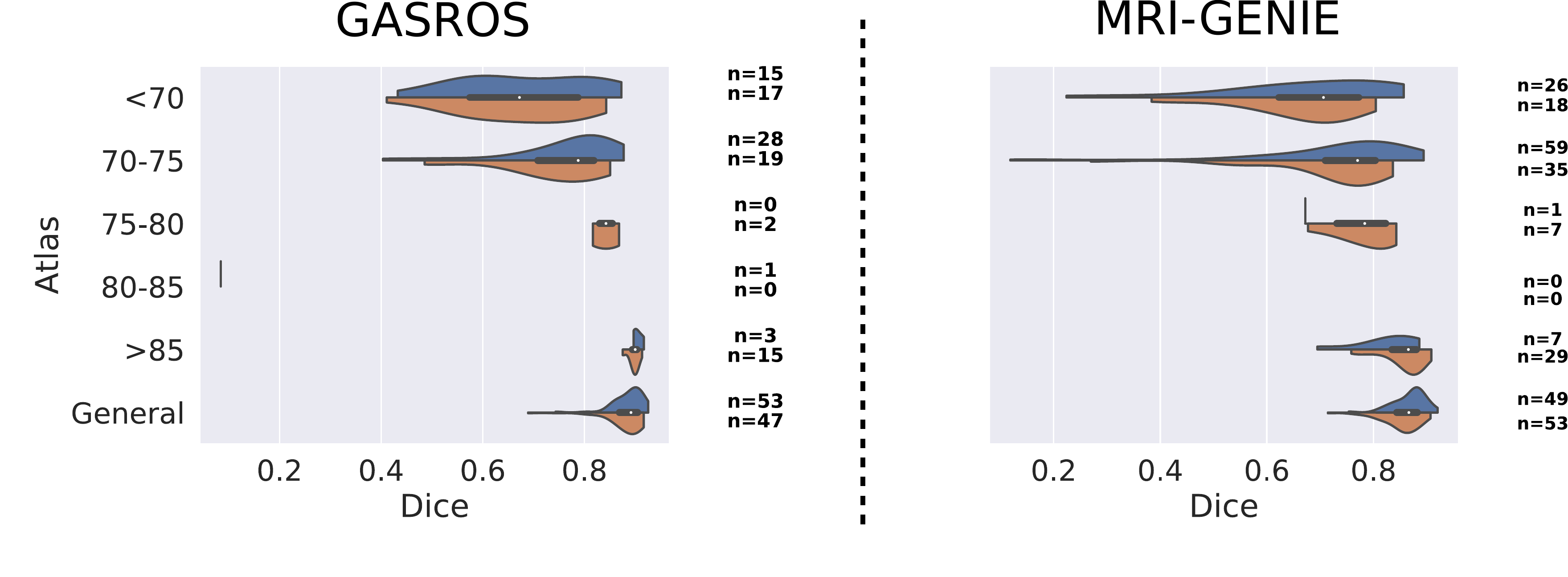}
\caption{\textbf{Comparison of multi-atlas registration using automated (blue) and manual (orange) segmentation of the ventricles in subject space.} The number of scans assigned to each atlas is indicated on the right of each plot for both automated and manual ventricle segmentations.}
\label{fig:ventSegManualAutomatedPlot}
\end{figure*}

To assess the validity of using the CNN results as reference for the registration, we evaluated the difference of results for the MAR framework in each dataset when using manually versus automatically segmented ventricles and found no large difference (Figure \ref{fig:ventSegManualAutomatedPlot} and Table \ref{table:ventSegManualAutomatedTable}).

\subsubsection{Spatial WMH maps of WMH burden}
\begin{figure*}[!t]
\centering
\includegraphics[height=13cm]{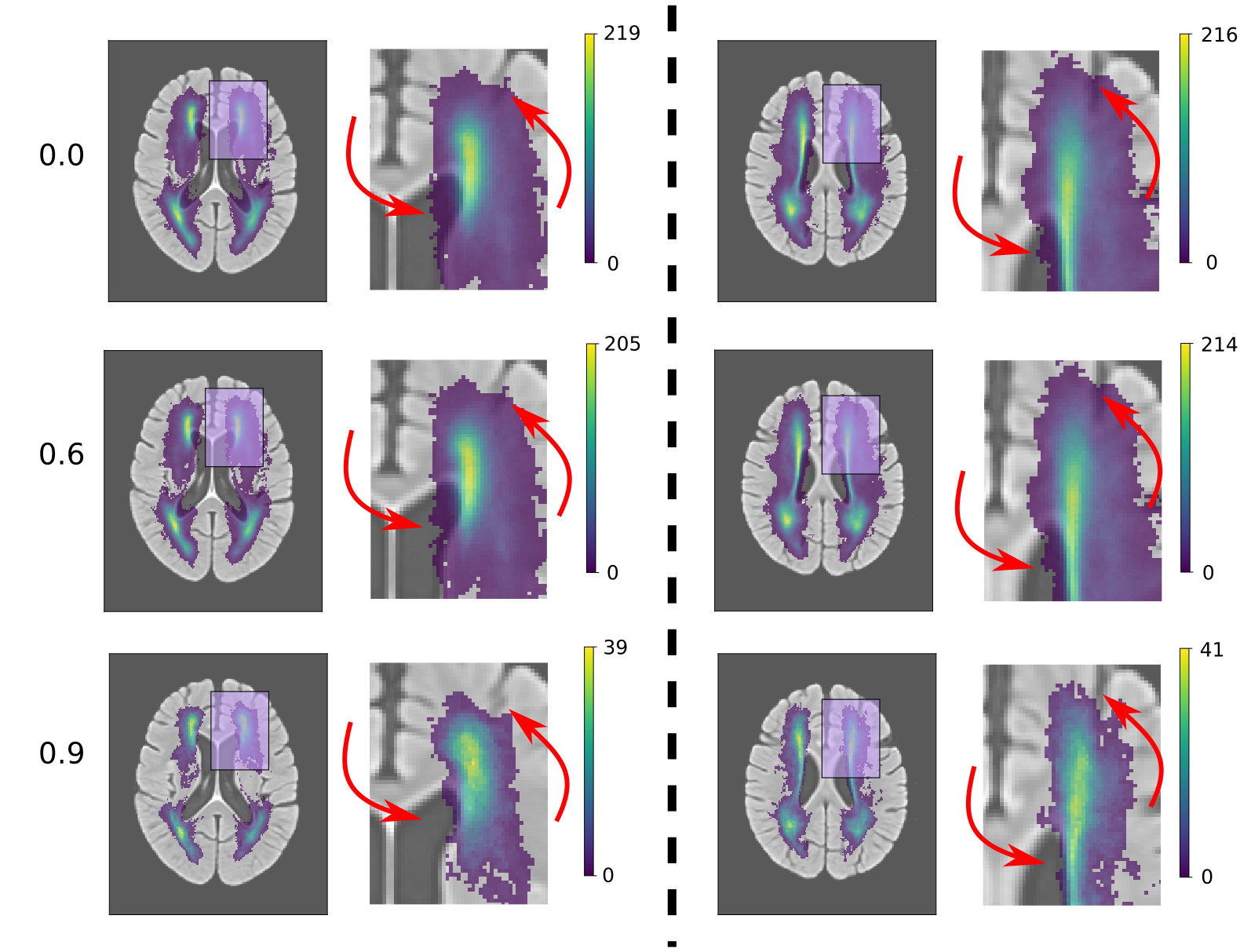}
\caption{\textbf{White matter hyperintensity (WMH) burden overlayed with the general atlas.} Rows correspond to different thresholds T for the quality of the registration measure $Q$ used to create WMH maps: from top to bottom: $Q \geq 0$ (all images = 791 images), $Q>0.6$ (748 images), and $Q>0.9$ (83 images). The columns correspond to two different brain slices in the axial plane. On the left of each column is the full image and on the right a zoomed in version of the region highlighted in pink. Red arrows indicate regions with a visible improvement in WMH maps.}
\label{fig:wmh}
\end{figure*}

Figure \ref{fig:wmh} shows that increasing the threshold of registration quality (rejecting more subjects) reduces, e.g., the erroneous extension of the WMH into the CSF compartments of the brain. 

\begin{table*}[!t]
\caption{\textbf{Gain in registration performance comparing the proposed multi-atlas registration framework using either the manual or automated ventricle segmentations to compute the registration quality.} Results are displayed as mean Dice coefficient of ventricle overlap. The number of scans assigned to each age-specific atlas is indicated between brackets.}
\resizebox{\textwidth}{!}{\begin{tabular}{l|l|l|l|l}
                                & GASROS 100 manual & GASROS 100 automated & MRI-GENIE manual & MRI-GENIE automated \\ \hline
Mean gain dice                  & 0.01 (100)        & 0.011 (100)          & 0.010 (142)      & 0.014 (142)         \\ \hline
Mean gain dice when improvement & 0.019 (53)        & 0.024 (43)           & 0.016(90)        & 0.022 (93)          \\ \hline
Under70                         & 0.034 (17)        & 0.04 (15)            & 0.036 (18)       & 0.033 (26)          \\ \hline
70-75                           & 0.019 (19)        & 0.019 (28)           & 0.016 (35)       & 0.019 (59)          \\ \hline
75-80                           & 0.001 (2)         & (0)’                & 0.004 (8)        & 0.0005 (1)          \\ \hline
80-85                           & (0)’             & 0.001 (1)            & (0)’            & (0)’               \\ \hline
Above 85                        & 0.005 (15)        & 0.006 (3)            & 0.006 (29)       & 0.008 (7)  
\label{table:ventSegManualAutomatedTable}        
\end{tabular}}
\end{table*}

\section{Discussion}
In this paper, we demonstrated the use of a ventricle segmentation algorithm using clinical FLAIR sequences, for automated registration quality assessment, and validated the proposed quality assessment metric in a multi-atlas registration (MAR) framework. 

The registration quality assessment method compared the ventricles of a subject, segmented with a machine learning algorithm, to the ventricles of the atlas, propagated to subject space. A ventricle segmentation algorithm that is robust to variations in scanners, sites and image resolutions is consequently a keypoint of its applicability. Here, we demonstrated that the proposed algorithm performed well in a multi-site scenario, while being trained with data from a single site. While, as expected, the algorithm reached a higher performance for the dataset it was optimized on (GASROS), the performance dropped by less than 6 percentage points of Dice coefficient when used on multi-site data. Importantly, the segmentation method generalized well to the other, multi-site data by designing appropriate data augmentation procedures, and without employing advanced transfer learning algorithms. Using manually or automatically segmented ventricles using the proposed deep learning algorithm, led to similar results with the MAR framework in each dataset (\ref{fig:ventSegManualAutomatedPlot} and \ref{table:ventSegManualAutomatedTable}), with a difference in mean gain in Dice coefficient of 0.001 in GASROS dataset, and 0.004 in the MRI-GENIE dataset. The largest differences were that: (1) when using the automated segmentation, more scans were assigned to atlas of age range 70-75 instead of atlas under 70 or the general atlas, and (2) when using the manual segmentation, more scans were assigned to atlas of age range above 85 instead of the general atlas. 

\cite{klein2009} showed that for multiple registration algorithms (including ANTS) the registration error of the ventricles correlates with registration errors in other regions. Manually annotated landmarks describing brain structures in the atlas could help to monitor more globally the registration quality than using the ventricles alone. However, automatically detecting such landmarks in clinical data remains a difficult task, and might lead to more erroneous cases, in contrast to segmenting a large, reliable structure, such as the ventricles. However, our framework can be extended to use multiple segmentations such as grey and white matter segmentations in the future.

We used the automated registration quality assessment method to design a multi-atlas registration (MAR) framework for improving registration quality. Instead of being directly and only registered to a general atlas, scans were first registered to atlases corresponding to several age categories. The best of these atlases was then chosen using the registration quality assessment method, and registration to the selected atlas was used as an intermediary registration step. In our dataset, using the MAR framework with ventricle overlap significantly improved the registration quality. Patients were often assigned to an intermediate atlas that was closer to their chronological age. However, we observed a shift, where, on average, subjects were matched to age-specific atlases of an older age category than their chronological age. This most probably resulted from the specific cohort in our analyses: all subjects had a prior acute ischemic event, which may reflect brains with increased biological age. This is further supported by studies which suggested that biological age, in contrast to chronological age, can play a key role in susceptibility to disease \citep{wang2019}.This suggests also that selecting the age-specific atlas using the patient’s chronological age would be a suboptimal strategy.

We further observed a positive correlation between ventricle volumes and the age category of the atlas the scans were assigned to. This relationship was expected, considering that age is positively correlated with ventricles volume in the general population \citep{walhovd2011}, which can also be seen on the age-specific atlases themselves (Figure \ref{fig:atlases}). The age-specific atlases also showed expected behavior of increased WMH volume and cortical atrophy with increasing age \citep{earnest1979}. In all experiments, only a few scans were assigned to the atlases of age category 75-80 and 80-85. Computing the mean squared intensity difference between the age-specific atlases and the general atlas revealed that atlas 75-80 was the closest to the general atlas, and atlas 80-85 was the most dissimilar. Consequently, scans most similar to atlas 75-80 were more likely to be assigned to the general atlas instead.

Other researchers have successfully used age-specific atlases \citep{sanchez2012,fillmore2015,liang2015,schirmer2019spatial, schirmer2019white}. \cite{liang2015} proposed to construct age-specific templates, and observed an improvement for hippocampi segmentation. And Fillmore et al. \cite{fillmore2015} observed an improvement in segmentation of white matter, gray matter and cerebrospinal fluid using an age-appropriate brain template. It is often impossible to find a single atlas, which works best for studies across the entire lifespan. Instead, using multiple age-specific atlases allows a more accurate description of the lifespan and can improve registration quality. In this article, we utilized five age groups, which already demonstrated improvement in overall registration quality. By using even more atlases, i.e. additional or smaller spaced age groups, could lead to further improvements.
Intermediary registration to a template has also been used to accelerate multi-atlas segmentation \citep{dewey2017}, or to improve registration from one image modality to another. For instance, \cite{parthasarathy2011} used a full-volume ultrasound image as intermediary image for the registration of live-3D ultrasound to MRI. Later, \cite{roy2014} used an synthesized CT image as intermediary image for the registration from MRI to CT.
Groupwise registration \citep{joshi2004,fletcher2009} could be another strategy to register all scans of a dataset to the same space. No template image needs to be selected in advance, and transformation fields are estimated simultaneously for all scans. One of the main disadvantages of groupwise registration is that the initial common space is estimated as the mean of all scans in the dataset. This mean image can be fuzzy and not provide enough guidance for the iterative optimization process \citep{wu2010}. Aligning the images to the MNI template instead of only aligning them to the general atlas created from ADNI healthy controls might be of interest, for example, to compare with other datasets already registered to the MNI template. For this purpose, a registration step to MNI template could be added as a last step of the MAR framework, after the registration to the general atlas. The general atlas would then need to be registered to the MNI template. This approach would guarantee a smoother and more controlled transformation than registering the age-specific atlases directly to the MNI template, and would provide a more precise monitoring of potential registration errors: ventricle overlap could be computed both when registering to the general atlas and when subsequently registering the MNI template, and errors in the pipeline could be more easily identified.

The proposed MAR framework using ventricle overlap could be categorized as a feature-based registration method. Segmentations in feature-based registration methods have already been used as initialization \citep{vemuri2003}, or have been optimized jointly with an intensity similarity metric for registration \citep{yezzi2003,pohl2006,chen2010}. More recently, \cite{balakrishnan2019} proposed to use a deep learning registration approach where segmentations of anatomical structures can be used as auxiliary data during the optimization. This would allow to include the ventricle segmentation in the optimization of the registration, instead of the proposed MAR framework. However, to date, utilizing auxiliary data for registration has not been tested in clinical scans, which are known to be substantially more challenging to segment and register. With the presented ventricle segmentation, and the segmentation of other structures and the entire brain, the extension of such approaches to clinical scans becomes more feasible and is of key interest for future studies. In Appendix \ref{appendix:auxiliary}, we compared a registration method in which ventricle segmentation was added as auxiliary objective with equal weight during registration to the proposed MAR and, as expected, obtained higher ventricles overlap. However, by utilizing the ventricle segmentation for registration, we cannot utilize it anymore for objectively assessing registration quality. Additionally, \cite{balakrishnan2019} have done similar experiments with brain registration and observed that when using the overlap of a single structure as auxiliary objective, the overlap of the other brain structures stayed either the same or even decreased when using larger weight for the auxiliary objective. In addition to, or instead of, using the ventricles to assess registration quality, it might also be interesting to inspect subcortical structures on T1-weighted MRI sequence, and attempt to exploit features based on the intensity difference between white and gray matter in, for example, the basal ganglia.
  
In our application, we demonstrated that it becomes feasible to automatically select only scans with high registration quality, leading to more globally accurate -- but also possibly more noisy as computed from a smaller set --  maps of WMH burden. Using automated assessment of registration quality to compute more accurate spatial patterns of disease could further help to relate spatial information to global phenotypes such as stroke severity or hypertension. For instance research has been done on how WMH distribution differs between patients with lobar intracerebral hemorrhage and healthy elderly \citep{zhu2012}, or on differences between deep and periventricular WMH in relation to stroke \citep{buyck2009}. However, discarding scans with a lower registration quality might also introduce a bias if the quality of the registration is related to one of the studied determinants or outcomes. Alternatively, a more rigorous quality control procedure might also be triggered for those scans.

There are limitations to this study. Our proposed method requires reliable automated segmentation of a key structure in the image, which can subsequently serves as a reference. This can be challenging with smaller structures in the image. Here, we focused on the ventricular system, which represents a structure that is relatively easy to segment consistently across subjects. While such a discriminative structure might not appear in every body part or with every imaging modality, further methodological advances in image segmentation will improve the generalizability of the proposed framework. Examples of structures that are suited to the proposed method could be large blood vessels in magnetic resonance angiography, or fetus in fetal MRI. The premise of our registration quality assessment lies in ventricles being visible on the clinical images. In particular in stroke cases, mass effects can alter the appearance of the ventricles, sometimes rendering the lateral ventricles invisible in the image. Additionally, the posterior horns of the ventricles may be masked due to the low resolution of the acquired clinical scans. If ventricles cannot be identified on the image, our proposed metrics may indicate insufficient registration quality. However, this assessment can be used to flag this subset of the registered scans as potentially erroneous, which can then be manually assessed by an expert rater rather than being completely rejected from the analysis. If the registration is erroneous, the third and fourth ventricles in particular are less likely to overlap with the atlas, reducing the probability of high dice for incorrect registration. We observed some outliers with low ventricle overlap between the automated and manual ventricle segmentation. The majority of these outliers -- for instance 2 out of 100 scans in GASROS dataset -- were scans with substantial motion artifacts, where the segmentation of ventricles was challenging even for human raters. Such scans are usually excluded from most neuroimaging pipelines. In addition, in some sites of the MRI-GENIE dataset, sulci were sometimes misclassified as ventricles. 
Another limitation is that the proposed MAR framework also multiplies the computation time by the number of atlases used: in our case, the registration is six times longer. However, each registration can be run in parallel, and in cases where immediate results are not necessary, this approach can help improve registration quality. Additionally, with the recent development of deep-learning based registration frameworks \citep{balakrishnan2019}, time concerns may become negligible.

Instead of using segmentation to perform automated quality control of registration, \cite{robinson2019} proposed to use registration to perform quality control of segmentation. This assumes that the registration is more robust to the variations present in the dataset than the segmentation. Using segmentation to perform automated quality control of registration assumes the opposite. Whether segmentation or registration can be considered more robust depends on the region of interest, imaging modality, and image resolution. The full ventricular system in the brain has a complex shape with substantial inter-subject variability due to, for example, brain atrophy and/or pathological processes. This makes the registration difficult when the shape of subject’s ventricles deviate from the expected ventricle shape. Conversely, image intensity on FLAIR-weighted MRI is a substantially more discriminative feature than shape. The high contrast between intensities inside and outside the ventricles is present in all subjects, scanner and FLAIR protocols. Segmentation of the ventricular system can therefore be expected to be more robust than registration. In contrast the structures composing the heart, as seen on MRI, have a simple ovoid shape with similar image intensities, making registration approaches more reliable as a reference. The other key aspect is that registration of clinical scans to templates is difficult and remains an open research question. Registration could potentially be more reliable if we had a more homogeneous, high-resolution dataset such a the UK-biobank, as \cite{robinson2019} used in their analyses.

Strengths of our work include segmentation of the four ventricles in clinical scans evaluated in multi-center data and more than 1000 scans. We introduced a multi-atlas registration framework based on this segmentation algorithm, and employed it to compute more accurate maps of WMH burden.

No single registration tool, or set of registration parameters, will perform best on all types of image qualities or sequences. By implementing an automated registration assessment step in large scale image analyses, it becomes feasible to test multiple registration pipelines and select the registration with the best performance. This can increase the number successful registrations, and potentially increase the sample size of a study without the need for time intensive manual quality assessment. 

In this work, we demonstrated the utility of an automated tool for assessing image registration quality in clinical scans. Importantly, in addition to extracting an additional phenotype from clinical scans -- namely the ventricle volume -- this image quality assessment step can be implemented in large-scale, automated processing pipelines of clinical MRI data, increasing the utility of such pipelines and offering improved quality of subsequent analysis, ultimately assisting in the translation of such pipelines to the clinic.

\section{Acknowledgements}
This project has been in part supported by the Foundation “De Drie Lichten” in the Netherlands (F.D.); The Netherlands Organisation for Health Research and Development (ZonMw) Project 104003005 (M. de B.); the European Union’s Horizon 2020 research and innovation programme under the Marie Sklodowska-Curie grant agreement No 753896 (M.D.S.); the NIH-NINDS K23NS064052, R01NS082285, and R01NS086905 (N.S.R.); American Heart Association/Bugher Foundation Centers for Stroke Prevention Research and Deane Institute for Integrative Study of Atrial Fibrillation and Stroke (N.S.R.). We gratefully acknowledge the support of NVIDIA Corporation with the donation of the Titan V GPU used for this research. 

Part of the Data collection and sharing for this project was funded by the Alzheimer's Disease Neuroimaging Initiative (ADNI) (National Institutes of Health Grant U01 AG024904) and DOD ADNI (Department of Defense award number W81XWH-12-2-0012). ADNI is funded by the National Institute on Aging, the National Institute of Biomedical Imaging and Bioengineering, and through generous contributions from the following: AbbVie, Alzheimer’s Association; Alzheimer’s Drug Discovery Foundation; Araclon Biotech; BioClinica, Inc.; Biogen; Bristol-Myers Squibb Company; CereSpir, Inc.; Cogstate; Eisai Inc.; Elan Pharmaceuticals, Inc.; Eli Lilly and Company; EuroImmun; F. Hoffmann-La Roche Ltd and its affiliated company Genentech, Inc.; Fujirebio; GE Healthcare; IXICO Ltd.; Janssen Alzheimer Immunotherapy Research and Development, LLC.; Johnson and Johnson Pharmaceutical Research and Development LLC.; Lumosity; Lundbeck; Merck and Co., Inc.; Meso Scale Diagnostics, LLC.; NeuroRx Research; Neurotrack Technologies; Novartis Pharmaceuticals Corporation; Pfizer Inc.; Piramal Imaging; Servier; Takeda Pharmaceutical Company; and Transition Therapeutics. The Canadian Institutes of Health Research is providing funds to support ADNI clinical sites in Canada. Private sector contributions are facilitated by the Foundation for the National Institutes of Health (www.fnih.org). The grantee organization is the Northern California Institute for Research and Education, and the study is coordinated by the Alzheimer’s Therapeutic Research Institute at the University of Southern California. ADNI data are disseminated by the Laboratory for Neuro Imaging at the University of Southern California.

\bibliography{biblio}


\cleardoublepage
\onecolumn
\appendix
\section{\textbf{Ventricle segmentation results for the 12 sites of MRI-GENIE.} The reported metrics are Dice coefficient (Dice), Jaccard index (Jaccard), true positive rate (TPR), volumetric similarity (VS), Mutual information (MI), Adjusted Rand Index (ARI), intraclass correlation coefficient (ICC), probabilistic distance (PBD), Cohen's kappa (KAP), Detection Error Rate (DER) and Outline Error Rate (OER).}
\label{appendix:ventSeg}

\begin{figure*}[!b]
\centering
\includegraphics[height=20cm]{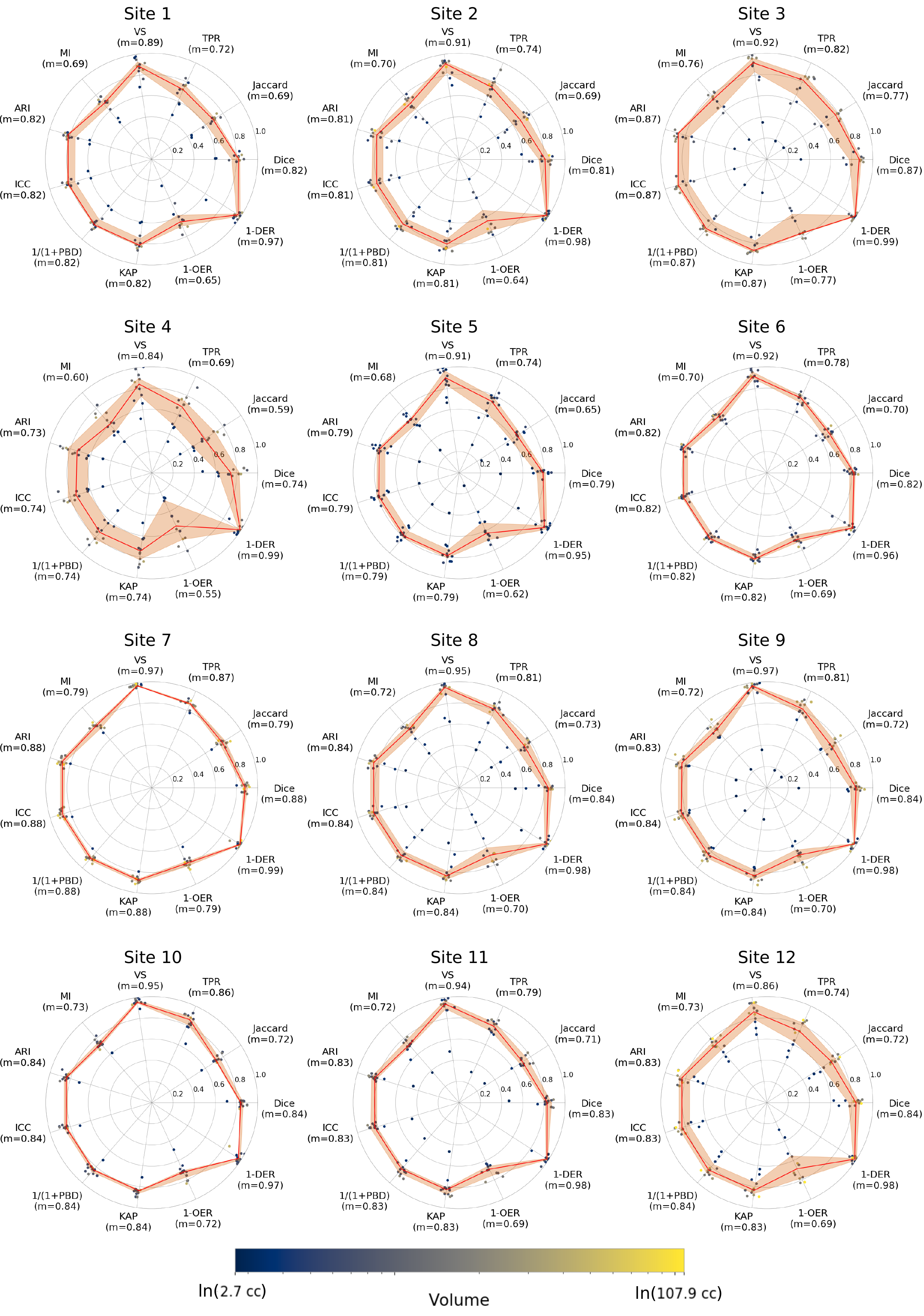}
\end{figure*}

\cleardoublepage
\section{List of ADNI 3 IDs used for the computation of the age-specific atlases.}
\label{appendix:ADNIdata}

$<70:$
\smallbreak
\verb|23_S_4448, 128_S_4607, 141_S_6008, 014_S_6076, 007_S_6120, 029_S_4384|
\bigbreak
$70-75:$
\smallbreak
\verb|068_S_4340, 031_S_4021, 094_S_4649, 003_S_4644, 019_S_4367, 070_S_4856, 009_S_4388, 100_S_4469,|
\indent\verb|135_S_4446, 068_S_4424, 116_S_4453, 016_S_4952, 137_S_4520, 127_S_4604, 037_S_4028, 129_S_4369,|
\indent\verb|014_S_4401, 135_S_6104, 029_S_4585, 037_S_4410, 024_S_4084, 135_S_4598|
\bigbreak
$75-80:$
\smallbreak
\verb|127_S_4148, 011_S_4105, 002_S_4225, 099_S_6038, 016_S_4951, 099_S_4076, 006_S_4357, 014_S_4576,|
\indent\verb|037_S_4308, 002_S_6007, 023_S_4164, 032_S_4277, 021_S_4335, 018_S_4400, 041_S_4427, 003_S_4288,|
\indent\verb|129_S_4422, 098_S_4275, 098_S_4506, 116_S_4483, 007_S_4488, 021_S_4276, 006_S_4485, 082_S_4428,|
\indent\verb|098_S_4003, 941_S_4292, 013_S_4580, 035_S_4464, 007_S_4637, 141_S_6061, 041_S_4200|
\bigbreak
$80-85:$
\smallbreak
\verb|041_S_4037, 011_S_4278, 127_S_0259, 068_S_0473, 018_S_4313, 019_S_4835, 002_S_1280, 032_S_0677,|
\indent\verb|006_S_0498, 067_S_0056, 007_S_4387, 070_S_5040, 068_S_0210, 141_S_0767, 007_S_1222, 123_S_0106|
\indent\verb|032_S_4429, 005_S_0602, 130_S_0969, 082_S_4224, 009_S_0751, 033_S_0734, 002_S_4213, 068_S_0127|
\indent\verb|002_S_1261, 027_S_0120, 137_S_4482, 067_S_0059, 006_S_0731, 033_S_1098, 941_S_4100, 123_S_0072|
\indent\verb|007_S_4620, 032_S_1169, 128_S_0272, 129_S_4396, 018_S_4399, 941_S_4376, 011_S_0021|
\bigbreak
$>85:$
\smallbreak
\verb|100_S_1286, 037_S_0303, 033_S_4177, 941_S_1195, 114_S_0416, 023_S_0031, 130_S_4343, 037_S_4071,|
\indent\verb|036_S_4491, 036_S_4389, 021_S_0337, 116_S_0382, 005_S_0610, 035_S_0156, 137_S_4466, 037_S_0454|
\indent\verb|123_S_0298, 099_S_4086, 033_S_1016, 941_S_4365, 033_S_4176, 126_S_0605, 002_S_0413, 126_S_0680|
\indent\verb|035_S_0555, 116_S_4855, 098_S_0896, 116_S_4043, 033_S_4179, 100_S_0069, 023_S_1190, 021_S_4254|

\section{Gain in registration performance by using the proposed multi-atlas registration method with ventricles overlap instead of the more standard cross-correlation for atlas selection. Left: GASROS. Right: MRI-GENIE. The registration quality with the proposed multi-atlas registration method $Q_{x,b}$ is in green; the registration quality with the proposed multi-atlas registration method using cross-correlation instead ventricle Dice to select the best atlas $Q_{x,bcc}$ is in pink; the overlay of both is purple. $\Delta_{b,bcc} = Q_{x,b} - Q_{x,bcc}$ , the gain in registration quality by using the proposed multi-atlas registration method with ventricles overlap instead of cross-correlation for the selection of the intermediary atlas is in blue.}
\label{appendix:gainCCplot}

\begin{figure*}[!htb]
\centering
\includegraphics[height=7cm]{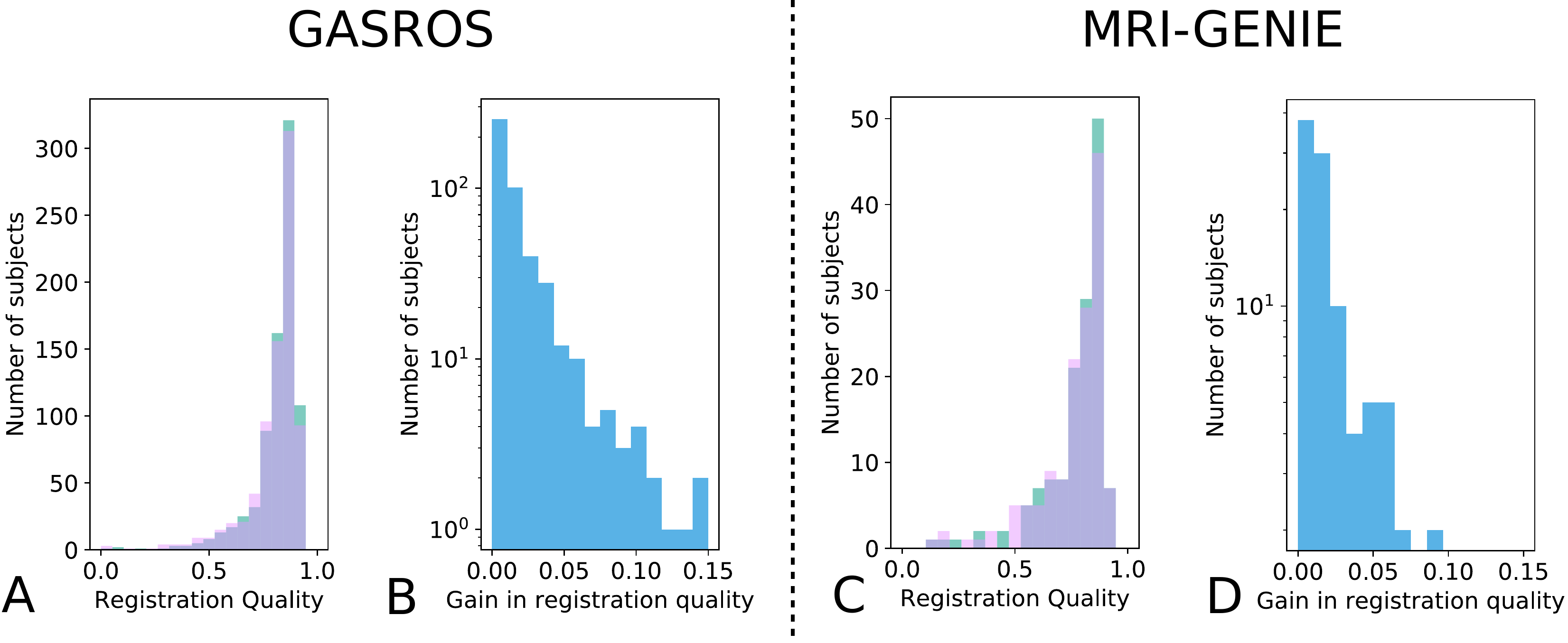}
\end{figure*}

\section{Gain in registration performance $\Delta_{b,bcc} = Q_{x,b} - Q_{x,bcc}$. Sample size is indicated in brackets.}
\label{appendix:gainCCtable}

\begin{table}[!htb]
\centering
\begin{tabular}{l|l|l}
                                & GASROS      & \multicolumn{1}{l}{MRI-GENIE} \\ \hline
Mean gain dice                  & 0.011 (791) & 0.014 (142)                    \\ \hline
Mean gain dice when improvement & 0.018 (468) & 0.021 (98)                    
\end{tabular}
\end{table}

\section{Gain in registration performance $\Delta_{b,g}$. Sample size is indicated in brackets.}
\label{appendix:gainMAR}

\begin{table}[!htb]
\centering
\begin{tabular}{l|l|l}
                                & GASROS      & \multicolumn{1}{l}{MRI-GENIE} \\ \hline
Mean gain dice                  & 0.012 (791) & 0.014 (142)                    \\ \hline
Mean gain dice when improvement & 0.022 (430) & 0.022 (93)                     \\ \hline
Under70                         & 0.038 (128) & 0.033 (26)                     \\ \hline
70-75                           & 0.014 (275) & 0.019 (59)                     \\ \hline
75-80                           & 0.006 (9)   & 0.0005 (1)                     \\ \hline
80-85                           & 0.14 (1)    & '(0)’                          \\ \hline
Above 85                        & 0.019 (17)  & 0.008 (7)                     
\end{tabular}
\end{table}

\section{{\color{mygreen}Dice score of ventricle overlap with direct registration to the general atlas (x-axis) vers registration with the proposed multi-atlas framework (y-axis) in GASROS and MRI-GENIE datasets.}}
\label{appendix:scatterplotDICE}

\begin{figure*}[!htb]
\centering
\includegraphics[height=5cm]{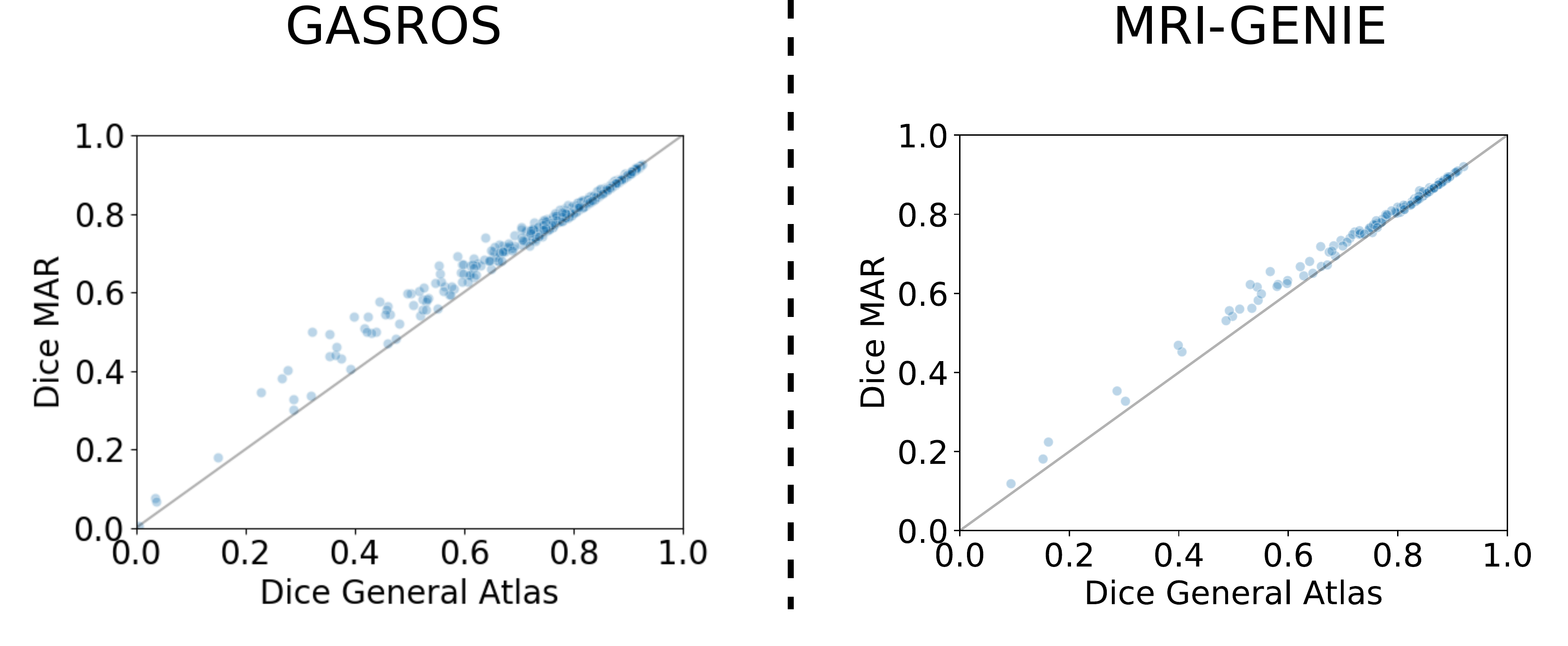}
\end{figure*}

\cleardoublepage
\section{Comparison of registration to the general atlas in which ventricle segmentation was added as auxiliary objective with equal weight during registration (GR -- guided registration -- left) to the proposed MAR (right). The value on the y-axis is the overlap between ventricles of the general atlas propagated to subject space and the ventricles segmented in subject space. **** indicates a p-value lower than 0.0001 for the Wilcoxon test}
\label{appendix:auxiliary}

\begin{figure*}[!htb]
\centering
\includegraphics[height=8cm]{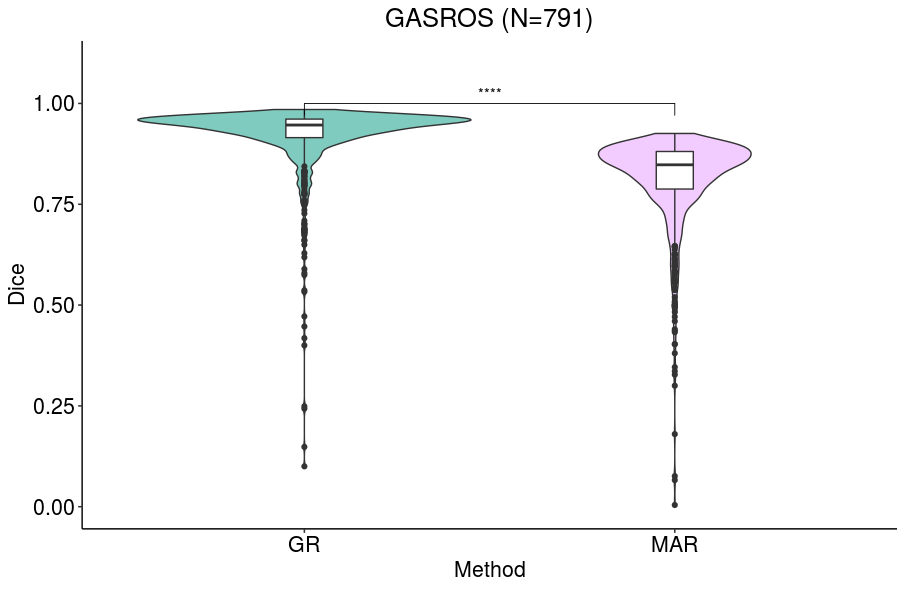}
\end{figure*}

\end{document}